\documentclass[12pt,psfig]{article}
\usepackage{amssymb}
\usepackage{amsmath}
\usepackage{epsfig}
\makeatletter
\def\@cite#1#2{{[{#1}]\if@tempswa\typeout
{IJCGA warning: optional citation argument
ignored: `#2'} \fi}}

\newcount\@tempcntc
\def\@citex[#1]#2{\if@filesw\immediate\write\@auxout{\string\citation{#2}}\fi
  \@tempcnta\z@\@tempcntb\m@ne\def\@citea{}\@cite{\@for\@citeb:=#2\do
    {\@ifundefined
       {b@\@citeb}{\@citeo\@tempcntb\m@ne\@citea\def\@citea{,}{\bf ?}\@warning
       {Citation `\@citeb' on page \thepage \space undefined}}%
    {\setbox\z@\hbox{\global\@tempcntc0\csname b@\@citeb\endcsname\relax}%
     \ifnum\@tempcntc=\z@ \@citeo\@tempcntb\m@ne
       \@citea\def\@citea{,}\hbox{\csname b@\@citeb\endcsname}%
     \else
      \advance\@tempcntb\@ne
      \ifnum\@tempcntb=\@tempcntc
      \else\advance\@tempcntb\m@ne\@citeo
      \@tempcnta\@tempcntc\@tempcntb\@tempcntc\fi\fi}}\@citeo}{#1}}
\def\@citeo{\ifnum\@tempcnta>\@tempcntb\else\@citea\def\@citea{,}%
  \ifnum\@tempcnta=\@tempcntb\the\@tempcnta\else
   {\advance\@tempcnta\@ne\ifnum\@tempcnta=\@tempcntb \else
\def\@citea{--}\fi
    \advance\@tempcnta\m@ne\the\@tempcnta\@citea\the\@tempcntb}\fi\fi}
\makeatother

\newcommand{\gsim}{\lower.7ex\hbox{$\;\stackrel{\textstyle>}{\sim}\;$}}
\newcommand{\lsim}{\lower.7ex\hbox{$\;\stackrel{\textstyle<}{\sim}\;$}}
\newcommand{\be}{\begin{equation}}
\newcommand{\ee}{\end{equation}}
\newcommand{\bea}{\begin{eqnarray}}
\newcommand{\eea}{\end{eqnarray}}

\usepackage{amssymb}

\def\baselinestretch{1}
\begin{document}
\catcode`@=11
\newtoks\@stequation
\def\subequations{\refstepcounter{equation}%
\edef\@savedequation{\the\c@equation}%
  \@stequation=\expandafter{\theequation}
  \edef\@savedtheequation{\the\@stequation}
  \edef\oldtheequation{\theequation}%
  \setcounter{equation}{0}%
  \def\theequation{\oldtheequation\alph{equation}}}
\def\endsubequations{\setcounter{equation}{\@savedequation}%
  \@stequation=\expandafter{\@savedtheequation}%
  \edef\theequation{\the\@stequation}\global\@ignoretrue

\noindent}
\catcode`@=12
\begin{titlepage}
\title{{\bf On the Newtonian limit of Generalized Modified Gravity Models}}
\vskip2in
\author{
{\bf Ignacio Navarro$$\footnote{\baselineskip=16pt E-mail: {\tt
ignacio.navarro@durham.ac.uk}}} $\;\;$and$\;\;$ {\bf Karel Van
Acoleyen$$\footnote{\baselineskip=16pt E-mail: {\tt
karel.van-acoleyen@durham.ac.uk}}}
\hspace{3cm}\\
 $$ {\small IPPP, University of Durham, DH1 3LE Durham, UK}.
}

\date{}
\maketitle
\def\baselinestretch{1.15}
\begin{abstract}
\noindent
We consider the Newtonian limit of modified theories of gravity that include inverse powers of the curvature in the action in order to explain the cosmic acceleration. It has been shown that the simplest models of this kind are in conflict with observations at the solar system level. In this letter we point out that when one adds to the action inverse powers of curvature invariants that do not vanish for the Schwarzschild geometry one generically recovers an acceptable Newtonian limit at small distances. Gravity is however modified at large distances. We compute the first correction to the Newtonian potential in a quite general class of models. The characteristic distance entering in these modifications is of the order of 10pc for the Sun and of the order of $10^2$ kpc for a galaxy.

\end{abstract}

\thispagestyle{empty} \vspace{5cm}  \leftline{}

\vskip-20.5cm \rightline{} \rightline{IPPP/05/27}
\rightline{DCPT/05/54} \vskip3in

\end{titlepage}
\setcounter{footnote}{0} \setcounter{page}{1}
\newpage
\baselineskip=20pt





It has been suggested that the origin of the accelerated expansion of the universe might be a modification of the dynamics of gravity at low curvatures, as opposed to a cosmological constant, or to a cosmologically evolving scalar field (quintessence). In \cite{Capozziello:2003tk,Carroll:2003wy} a correction to the action of the form $\delta {\cal L}=- M_p^2 \mu^4/R$ was proposed. $M_p$ is the Planck mass, $R$ is the curvature scalar and $\mu$ is a parameter with dimensions of mass. It was shown that this model has cosmologically interesting solutions that can explain the acceleration of the universe without recourse to dark energy if $\mu \sim H_0$. However, it was soon realized that this model is in conflict with solar system experiments \cite{Chiba:2003ir} and presents instabilities when matter is present \cite{Dolgov:2003px}. Several modifications were suggested to revive this explanation of the cosmic acceleration, like adding positive powers of the curvature or taking into account quantum effects to cancel the instabilities \cite{Nojiri:2003ft,Nojiri:2003wx}, trading this correction for a singular $ln(R)$ term in the action \cite{Nojiri:2003ni} or working in the Palatini formalism, that yields second order field equations instead of fourth order ones \cite{Vollick:2003aw}. However one can generally expect problems in reproducing a correct Newtonian limit for models involving inverse powers of the curvature scalar (or Ricci tensor) only \cite{Dick:2003dw,Flanagan:2003rb}.

In this letter we point out that if we add to the action inverse powers of curvature invariants that do not vanish for the Schwarzschild solution (so they have to involve the Weyl or Riemann tensors) one can make an expansion of the exact solution over the Schwarzschild geometry and it is easy to see that at small distances the corrections are suppressed. This can be seen already at the level of the Newtonian action obtained by a weak field expansion of the relativistic one. Lets consider for definiteness a term in the action of the form $\delta {\cal L}= M_p^2 \mu^{4n+2}/Q^n$ where $Q\equiv R_{\mu\nu\sigma\lambda}R^{\mu\nu\sigma\lambda}$ and $n$ is some positive integer number (this kind of corrections have been considered in \cite{Carroll:2004de} in the context of cosmological evolution). In the Schwarzschild geometry, outside a spherical mass distribution, this invariant reads $Q=\frac{48(GM)^2}{r^6}$ where $M$ is the mass of the object sourcing the gravitational field and $r$ is the distance to the center of the object. We can then expect that the modifications induced by adding some $inverse$ powers of this term to the action will be suppressed at small distances for the same reason its effect on cosmology becomes important only at late times, when the mean curvature of the universe is $R\sim \mu^2$. To get an order of magnitude of the expected characteristic distance controlling the departure from Newtonian gravity we can evaluate the Newtonian action in this geometry (so we consider the weak field limit $r\gg GM$)
\be
S = \int d^4 x \left[(\nabla \phi)^2 + 8\pi G\phi \rho \right],
\ee
where $\phi \simeq -(1+g_{00})/2$ is the Newtonian potential, $\rho$ is the energy density and $\phi = -\frac{GM}{r}$ for a static point-like mass. We can now consider the shift in the action induced by the correction 
\be
\delta S = \int d^4 x \;\; \mu^{4n+2}/Q^n \simeq \int d^4 x \;\; \mu^{4n+2}\frac{r^{6n}}{48^n (GM)^{2n}},
\ee
when evaluated at the previous solution. We see that this correction will be negligible at small distances and will become important roughly when $\mu^{4n+2}\frac{r^{6n}}{48^n (GM)^{2n}} \sim(\nabla \phi)^2= \frac{(GM)^2}{r^4}$. This implies a characteristic scale for the induced modification of Newtonian gravity given by
\be
r_c^{6n+4} \equiv \frac{\left(GM\right)^{2n+2}}{\mu^{4n+2}}.\label{rc}
\ee
One might think that gravity is well measured at large distances, so this kind of modifications would be in conflict with observations, but due to the extremely small value of $\mu$ necessary to explain the cosmic acceleration ($\mu \sim H_0$) the corresponding distance at which this correction becomes important is astronomically large. For instance for a star like the Sun this distance is of order $\sim 10-100$ parsec, much bigger than the solar system, while for a typical galaxy like the milky way $r_c$ is of order $\sim 10^2-10^3$ kilo-parsec, with lower values corresponding to lower values of $n$ (we are taking $n \geq 1$). This last value is not far from the size of the galaxy itself, so this suggests that the modifications of gravity induced by these terms will be negligible at the level of solar system experiments but can have a significant impact in the dynamics of larger objects like a galaxy.

To confirm this intuition we will solve at first order the modified Einstein equations, for a class of models, in an expansion over the Schwarzschild geometry. We will obtain the first correction to the gravitational potential and we will check that the scale controlling the departure from Newtonian gravity at large distances is indeed given by the formula above. So we start with the action
\be S=\int \!\!d^4x\sqrt{-g}\frac{1}{16\pi G}\left[R+f(R,P,Q)\right]\,,\ee with  \bea P&\equiv&R_{\mu\nu}R^{\mu\nu}\,,  \eea and $Q$ defined before as the contraction of two Riemann tensors. 
We are using the notation of \cite{Carroll:2004de}, and we will also take the same form for $f(R,P,Q)$, \be f(R,P,Q)=-\frac{\mu^{4n+2}}{(aR^2+bP+cQ)^n}\,. \ee
With this choice for $f(R,P,Q)$, corrections to standard cosmology will only occur at the present epoch if we take $\mu\sim H_0$ and it has been shown that for appropriate values of the parameters there are cosmologically interesting accelerating solutions that are attractors at late times \cite{Carroll:2004de}.

Our strategy
will be then to take the ansatz \be ds^2=-[1-\frac{2GM}{r}+\epsilon
A(r)]dt^2+[1-\frac{2GM}{r}+\epsilon B(r)]^{-1}dr^2+r^2d\Omega^2_2\,,\ee
and treat $\epsilon$ as a small expansion parameter. We can write the
equations of motion as \be G_{\mu\nu}+\mu^{4n+2}H_{\mu\nu}=0\,,\ee where
$G_{\mu\nu}$ is the usual Einstein tensor and $\mu^{4n+2}H_{\mu\nu}$ is the
extra term generated by $f(R,P,Q)$. We expand now both tensors in $\epsilon$:
\bea G_{\mu\nu}&=& G_{\mu\nu}^{(0)}+\epsilon G_{\mu\nu}^{(1)}+\epsilon^2
G_{\mu\nu}^{(2)}+\ldots\,, \nonumber\,\\ H_{\mu\nu}&=&
H_{\mu\nu}^{(0)}+\epsilon H_{\mu\nu}^{(1)}+\epsilon^2
H_{\mu\nu}^{(2)}+\ldots\,.\eea Since the Schwarzschild solution solves the
ordinary Einstein equations, we have $G_{\mu\nu}^{(0)}=0$. Treating $\mu^{4n+2}$ as an order $\epsilon$ parameter, at first order
in our expansion the equations for $A$ and $B$ become (from now
on we set $\epsilon=1$) \be G_{\mu\nu}^{(1)}=-\mu^{4n+2}H_{\mu \nu}^{(0)}\,. \ee
For the $tt$ component of this equation we find: \bea
\frac{(2GM-r)(B+rB')}{r^3}&=&\left(\frac{\mu}{2}\right)^{2+4n}\frac{r^{6n+1}}{(GM)^{2n+1}}\frac{1}{(3c)^n}\left[(16n+40n^2+24n^3) \right. \nonumber \\  && \left.+\frac{GM}{r}(2-58n-156n^2-96n^3)\nonumber\right.\\&&\left.+\left(\frac{GM}{r}\right)^2(-4+52n+152n^2+96n^3)\right]\,,
\eea while the $rr$ component reads: \bea\frac{
-2GMA+rB+r(-2GM+r)A'}{r(r-2GM)^2}=\left(\frac{\mu}{2}\right)^{2+4n}\frac{r^{6n+2}}{(GM)^{2n+1}}\frac{(1+n)}{(3c)^n}\frac{\left[4n-\frac{GM}{r}(2+12n)\right]}{(r-2GM)}\,.\eea
Notice that $H_{\mu\nu}^{(0)}\rightarrow 0$ for $r\rightarrow 0$ (and so do $A$ and $B$) and we approach the Schwarzschild solution in this limit. Also it is clear that we can not take the limit $c\rightarrow 0$, since $H^{(0)}_{\mu\nu}$ diverges in this case and our expansion breaks down. In fact our equations are completely independent of $a$ and $b$ at the order we are working, since both $R$ and $P$ vanish in our zero-th order background. We can solve exactly the previous equations yielding 
\be
B(r) =- \left(\frac{\mu}{2}\right)^{4n+2}\frac{2(1+n)r^{6n+3}}{(3c)^n(6n+3)(GM)^{2n+1}}\left[(1-14n-24n^2)\frac{GM}{r}+6n(1+2n)\right], \ee
\be
A(r) =- \left(\frac{\mu}{2}\right)^{4n+2}\frac{2(1+n)r^{6n+3}}{(6n+3)(3c)^n(GM)^{2n+1}}\left[\frac{GM}{r}(6n+1)-4n\right].
\ee
This solution automatically solves the other equations because of the
Bianchi identities: $\Delta_{\mu}
G^{\mu\nu}=\Delta_{\mu} H^{\mu\nu}=0$. For $r\gg GM$ we can approximate
\be A(r)=\alpha\frac{GM}{r_c} \left(\frac{r}{r_c}\right)^{6n+3}\,\,\,;\,\,\,B(r)=-\alpha\frac{6n+3}{2} \frac{GM}{r_c}\left(\frac{r}{r_c}\right)^{6n+3}\,,\label{solution}\ee where
$\alpha\equiv \frac{8n(1+n)}{(6n+3)2^{4n+2}(3c)^n}$ and $r_c$ is defined in eq.(\ref{rc}). The correction to the Newtonian potential then reads: \be \phi(r) \simeq -\left[1 - \frac{\alpha}{2}\left(\frac{r}{r_c}\right)^{6n+4}\right]\frac{GM}{r} \label{potential}. \ee So we find indeed that the correction to Newtonian gravity becomes sizable at distances $r\sim r_c$. By construction, our expansion breaks down at this same distance. We can, however, trust our approximate solution when $r\ll r_c$. We can check the self-consistency of our expansion by inserting our approximate solution in the exact equation, for which we get that $G_\mu^\nu+\mu^{4n+2}H_\mu^\nu \sim (GM/r^3)(r/r_c)^{12n+8}$ while the terms we have solved for are of order $G_{\mu}^{\nu(1)}=-\mu^{4n+2}H_{\mu}^{\nu(0)} \sim (GM/r^3)(r/r_c)^{6n+4}$. In fact we can consider our approximate solution as the first order of the Taylor expansion in powers of $\mu^{4n+2}\propto 1/r_c^{6n+4}$, that is generated by solving the equations of motion, order by order. For $r\gg GM$ this expansion takes the form
\bea ds^2&\simeq &-\left[1-\frac{2GM}{r}\left( 1 - \frac{\alpha}{2} \left(\frac{r}{r_c}\right)^{6n+4}+ {\cal O}\left((r/r_c)^{12n+8}\right)\right)\right]dt^2  \\&+&\left[1-\frac{2GM}{r}\left( 1 + \frac{\alpha(6n+3)}{4} \left(\frac{r}{r_c}\right)^{6n+4}+ {\cal O}\left((r/r_c)^{12n+8}\right)\right)\right]^{-1}dr^2+r^2d\Omega^2_2\,,\nonumber\eea and the validity of the expansion is controlled by the parameter $r/r_c$.

The full equations of motion for the metric are now fourth order, so Birkhoff's Theorem does not hold and the solution for a spherically symmetric mass distribution will not be unique in general. However only first derivatives of the metric appear in the first order of our expansion, so the solution is unique at this order \footnote{In this case the two boundary conditions determine the mass and an overall normalization of the time coordinate.}. Also, we have considered an expansion over an asymptotically flat geometry. One might be worried because flat space is not a solution in our theory, and the equations actually become divergent for this geometry (as $r\rightarrow \infty$ in our zero-th order background). But considering the spacetime asymptotically flat or not (we could consider our expansion in the background of one of the cosmological solutions obtained in \cite{Carroll:2004de}, for instance) would only introduce appreciable corrections at distances much bigger than $r_c$,  where the curvature induced by the spherical mass distribution has fallen down to the cosmological value $Q\sim H_0^4$. And, as we said before, our expansion is only valid for distances $r\ll r_c$. We should mention here that our solution differs from the solution obtained in \cite{Chiba:2005nz} for the linearized version of the same theory over a maximally symmetric space. The reason is that the linearized expansion of the theory considered in \cite{Chiba:2005nz} breaks down for small radius in the spacetime of a spherically symmetric mass, where the curvature is substantially larger than that of the cosmic background.

It is remarkable that with the values of the parameters necessary to explain the cosmic acceleration ($\mu \sim H_0$) the correction (\ref{potential}) is negligible for solar system scales (within the solar system the relative corrections with respect to the Newton potential are smaller than $\sim 10^{-10-38n}$) but becomes important at galactic lengths. So despite having a negligible effect on the solar system this kind of modifications would have implications for the Dark Matter problem, changing the required abundances of Dark Matter and would in fact influence the measurement of most cosmological parameters. It is tempting to speculate on the construction of a MOND-like modification of gravity \cite{Milgrom:1983pn} along these lines that could explain the galactic rotation curves without recurring to Dark Matter. The model we have analyzed goes in that direction, in the sense that the gravitational attraction becomes stronger at long distances, but the size of the correction appears to be too small. Notice however that given the perturbative method we have used to solve the equations we can not say that much about the domain $r \sim r_c$ where the extra contribution to the gravitational force is of the order of the Newtonian one. We leave the issue of determining the nature of the corrections at these scales for future work. Also, we have not studied other aspects of the model like the possible appearance of ghosts \cite{Chiba:2005nz,Nunez:2004ts} and instabilities \cite{Dolgov:2003px}. These issues are more model dependent\footnote{As we said instabilities can be controlled for instance by the inclusion of higher order terms \cite{Nojiri:2003ft}, while the appearance of ghosts in an effective low energy action can be reconciled with observations if the theory is replaced by another at high energies \cite{Cline:2003gs}.}, and in this letter we wanted to focus on the general mechanism for reproducing a correct Newtonian limit at small distances in this class of models. But given that we are introducing a correction in the action that gets smaller as $r\rightarrow 0$ for a spherical mass, one might also expect that any pathological behaviour associated with this term will only manifest itself at large distances.

\section*{Acknowledgements}

We thank Jose Santiago for conversations and the Fund for Scientific Research Flanders (Belgium) and PPARC for financial support.

\end{document}